\documentclass[aps,prb,nobibnotes,twocolumn,superscriptaddress,bibliography]{revtex4-2}

\usepackage{amsfonts}
\usepackage{mathrsfs}
\usepackage{amsmath}% needed for subequations
\usepackage{color}
\usepackage{graphicx}
\usepackage{bm}% bold maths
\usepackage{amssymb}
\usepackage{xspace}
\usepackage{epstopdf}
\usepackage{dcolumn}% Align table columns on decimal point
\usepackage{longtable}
\usepackage{multirow}
\usepackage{float}
\usepackage{comment}
\usepackage[colorlinks=true, letterpaper=true, pdfstartview=FitV, linkcolor=blue, citecolor=blue, urlcolor=blue]{hyperref}

\makeatother

\begin{document}
\title{Charge-4 Weyl point: Minimum lattice model and chirality-dependent
properties }
\author{Chaoxi Cui}
\affiliation{Centre for Quantum Physics, Key Laboratory of Advanced Optoelectronic
Quantum Architecture and Measurement (MOE), School of Physics, Beijing
Institute of Technology, Beijing, 100081, China}
\affiliation{Beijing Key Lab of Nanophotonics \& Ultrafine Optoelectronic Systems,
School of Physics, Beijing Institute of Technology, Beijing, 100081,
China}
\author{Xiao-Ping Li}
\affiliation{Centre for Quantum Physics, Key Laboratory of Advanced Optoelectronic
Quantum Architecture and Measurement (MOE), School of Physics, Beijing
Institute of Technology, Beijing, 100081, China}
\affiliation{Beijing Key Lab of Nanophotonics \& Ultrafine Optoelectronic Systems,
School of Physics, Beijing Institute of Technology, Beijing, 100081,
China}
\author{Da-Shuai Ma}
\affiliation{Centre for Quantum Physics, Key Laboratory of Advanced Optoelectronic
Quantum Architecture and Measurement (MOE), School of Physics, Beijing
Institute of Technology, Beijing, 100081, China}
\affiliation{Beijing Key Lab of Nanophotonics \& Ultrafine Optoelectronic Systems,
School of Physics, Beijing Institute of Technology, Beijing, 100081,
China}
\author{Zhi-Ming Yu}
\email{zhiming\_yu@bit.edu.cn}

\affiliation{Centre for Quantum Physics, Key Laboratory of Advanced Optoelectronic
Quantum Architecture and Measurement (MOE), School of Physics, Beijing
Institute of Technology, Beijing, 100081, China}
\affiliation{Beijing Key Lab of Nanophotonics \& Ultrafine Optoelectronic Systems,
School of Physics, Beijing Institute of Technology, Beijing, 100081,
China}
\author{Yugui Yao}
%\email{ygyao@bit.edu.cn}

\affiliation{Centre for Quantum Physics, Key Laboratory of Advanced Optoelectronic
Quantum Architecture and Measurement (MOE), School of Physics, Beijing
Institute of Technology, Beijing, 100081, China}
\affiliation{Beijing Key Lab of Nanophotonics \& Ultrafine Optoelectronic Systems,
School of Physics, Beijing Institute of Technology, Beijing, 100081,
China}
\begin{abstract}
Topological Weyl semimetals have been attracting broad interest. Recently, a new type of Weyl point with topological charge of $4$, termed as charge-4 Weyl point (C-4 WP), was proposed in spinless systems. Here, we show the minimum symmetry requirement for C-4 WP is point group $T$ together with ${\cal T}$ symmetry or point group $O$. We establish a minimum tight-binding model for C-4 WP on a cubic lattice with time-reversal symmetry and without spin-orbit coupling effect. This lattice model is a two-band one, containing only one pair of C-4 WPs with opposite chirality around Fermi level. Based on both
the low-energy effective Hamiltonian and the minimum lattice model,
we investigate the electronic, optical and magnetic properties of
C-4 WP. Several chirality-dependent properties are revealed, such
as chiral Landau bands, quantized circular photogalvanic effect and quadruple-helicoid surface arc states. Furthermore, we predict that under
symmetry breaking, various exotic topological phases can evolve out
of C-4 WPs. Our work not only reveals several interesting phenomena
associate to C-4 WPs, but also provides a simple and ideal lattice
model of C-4 WP, which will be helpful for the subsequent study on C-4 WPs.
\end{abstract}
\maketitle

\section{Introduction }

Since the prediction of Weyl semimetal \cite{wan2011topological,murakami2007phase}, the search of emergent
particles in three-dimensional (3D) crystalline materials has been actively performed \cite{chiu2016classification,armitage2018weyl,zhang2019catalogue,vergniory2019complete,
tang2019comprehensive}. These emergent particles can produce many interesting phenomena that can not be found in conventional metals \cite{burkov2015chiral,yu2016predicted,lu2017quantum,chen2019weak,liu2020quantized}.
Various emergent particles have been predicted and some of them have been confirmed in experiments \cite{fang2012multi,yang2014classification,liu2014discovery,huang2015observation,yang2015weyl,weng2015topological,xu2015experimental,lv2015experimental,zhu2016triple,deng2016experimental,bradlyn2016beyond,gooth2017experimental, wu2018nodal,yu2019quadratic,wu2020higher,lv2021experimental}.
%(cite Weng hongming TaAs; NaBi3, nodal line in C-network material, weikang prb nodal surface, ziming PRX triple point, fang ceng prl multiple-weyl point, B. J. Yang NC 3D Dirac point, zhi-ming prb higher-order nodal line, weikang   prb higher-order DP and some experimental works published in Nature, Science, NP, PRX...)

Particularly, a complete classification of all the possible emergent particles in 3D systems with time-reversal symmetry (${\cal T}$) has been recently  presented by Yu \emph{et al.} \cite{yu2021encyclopedia}, which gives 19 types of spinless particles and 23 types of spinful particles. Among these emergent particles, the two-fold degenerate points, namely, the Weyl points (WPs), always attract the most concern, as they have the simplest models but nontrivial physics.

In Weyl semimetals, the WPs denote isolated doubly degenerate band crossings in the Brillouin zone (BZ), around which the low-energy excitation can be described by the Weyl equation rather than the conventional Schr$\ddot{o}$dinger equation. The conventional WP exhibits linear dispersion along any direction in momentum space and has a topological charge (Chern number) of ${\cal C}=\pm1$. Hence, the conventional WP also is termed as charge-1 WP (C-1 WP) \cite{yu2021encyclopedia}. Due to the nonzero topological charge, the WPs should come in pair with opposite chirality and their projections on boundary are connected by the Fermi arc \cite{wan2011topological,armitage2018weyl}. Besides, the topological charge can leads to many intriguing electronic, optical and magnetic phenomena \cite{nagaosa2020transport}, which generally have a strong dependence on the sign of the topological charge, e.g. the chirality. With multiple symmetries, two or three C-1 WPs with same topological charge may merge together, giving rise to the Charge-2 WP (C-2 WP) and Charge-3
Weyl point (C-3 WP) \cite{xu2011chern,fang2012multi}. In contrast to the C-1 WP, the C-2 (C-3) WP features quadratic (cubic) dispersion in a plane and a linear dispersion out-of-plane. Since the band dispersion would affect the shape of Fermi surface and then determines most of the material properties, the C-2 WP and
C-3 WP are demonstrated to exhibit signatures distinguished from the  C-1 WPs \cite{han2019emergent,wang2019topological,zhang2021quantum}.

Currently, the exploration of emergent particles is extended to spinless
systems, such as phonon spectrum and artificial photonic and phononic crystals \cite{stenull2016topological,li2017dirac,zhang2018topological,
imhof2018topolectrical,yang2018ideal,ozawa2019topological,
cooper2019topological,xiao2020experimental}.
The spinless systems are fundamentally different from the spinful systems, reflected in the absence of spin-orbit coupling (SOC) effect.
Thus, the time-reversal symmetry ${\cal T}$ satisfies ${\cal T}^{2}=1$
in spinless systems, while ${\cal T}^{2}=-1$ in spinful systems.
As a consequent, some emergent particles that can not appear in spinful
systems would emerge in spinless materials, and vice versa. Recently,
a new kind of Weyl point with topological charge of ${\cal C}=\pm4$,
e.g. C-4 WP, has been proposed by several independent groups in spinless
systems \cite{zhang2020twofold,yu2021encyclopedia,liu2020symmetry},
which has a cubic dispersion along $(111)$ direction and quadratic dispersion along any other direction. The C-4 WP was first predicted by Zhang et. al. in Ref. {\cite{zhang2020twofold}} (where the C-4 WP is termed as twofold quadruple Weyl node). They found that the C-4 WPs can appear in both the electronic band structure in the absence of SOC effect and the phonon spectra of a series of LaIrSi-type materials. Liu et al.\cite{liu2020symmetry}
also discovered this kind of WP when they systematically studied the symmetry-enforced WP in spinless systems. Yu et al. \cite{yu2021encyclopedia}, based on
the encyclopedia of emergent particles they established, concluded
that the C-4 WP is the last two-fold degenerate point in 3D crystals
and only appears in spinless systems.

Motivated by the discovery of the new emergent particle, in this work,
we study the minimum symmetry requirement of the C-4 WP, as well as
the chirality-dependence of the electronic, optical and magnetic properties. There are two different minimum symmetry conditions
that can protect the C-4 WP. One is the point group (PG) $O$ and
the other is the PG $T$ together with ${\cal T}$ symmetry. We then establish a minimum tight-binding (TB) model of the  C-4 WP on a cubic lattice with PG $O$ and ${\cal T}$ symmetry. This model has two bands and only one pair of C-4 WPs with opposite topological charge locating at $\Gamma$ ($000$) point and $R$ ($\pi\pi\pi$) point, respectively. Four extensive Fermi arcs connecting the projection of the two WPs appear on each boundary of the system. Particularly, the Fermi arcs form  unconventional quadruple-helicoid surface states around each WP, and the chirality of the quadruple-helicoid surface states is determined by the chirality of the WP.
We also investigate Landau level (LLs) and circular photogalvanic effect (CPGE) of the C-4 WPs, and find that both  have a strong dependence on the chirality of the C-4 WP.
The chirality-dependent properties of the C-4 WP are further demonstrated by the minimum lattice model.
At last, the possible phase transition of C-4 WP under symmetry descend are studied.
Our work not only predicts several novel properties of a newly found emergent particle but also establishes a minimum lattice model of
the C-4 WP, which would be helpful for the further investigation on this novel particle.

\section{Charge-4 Weyl point }

\subsection{Minimum symmetry requirement }

In the previous works, all the possible space groups that can protect
the C-4 WP have been presented \cite{zhang2020twofold,yu2021encyclopedia,liu2020symmetry}.
In this subsection, we study the symmetry
protection of the C-4 WP in detail. According to previous studies,
one knows that the minimum symmetry requirement for C-4 WP is PG $O$
or the PG $T$ together with ${\cal T}$ symmetry.

\begin{figure}
\begin{centering}
\includegraphics[width=8cm,height=8cm,keepaspectratio]{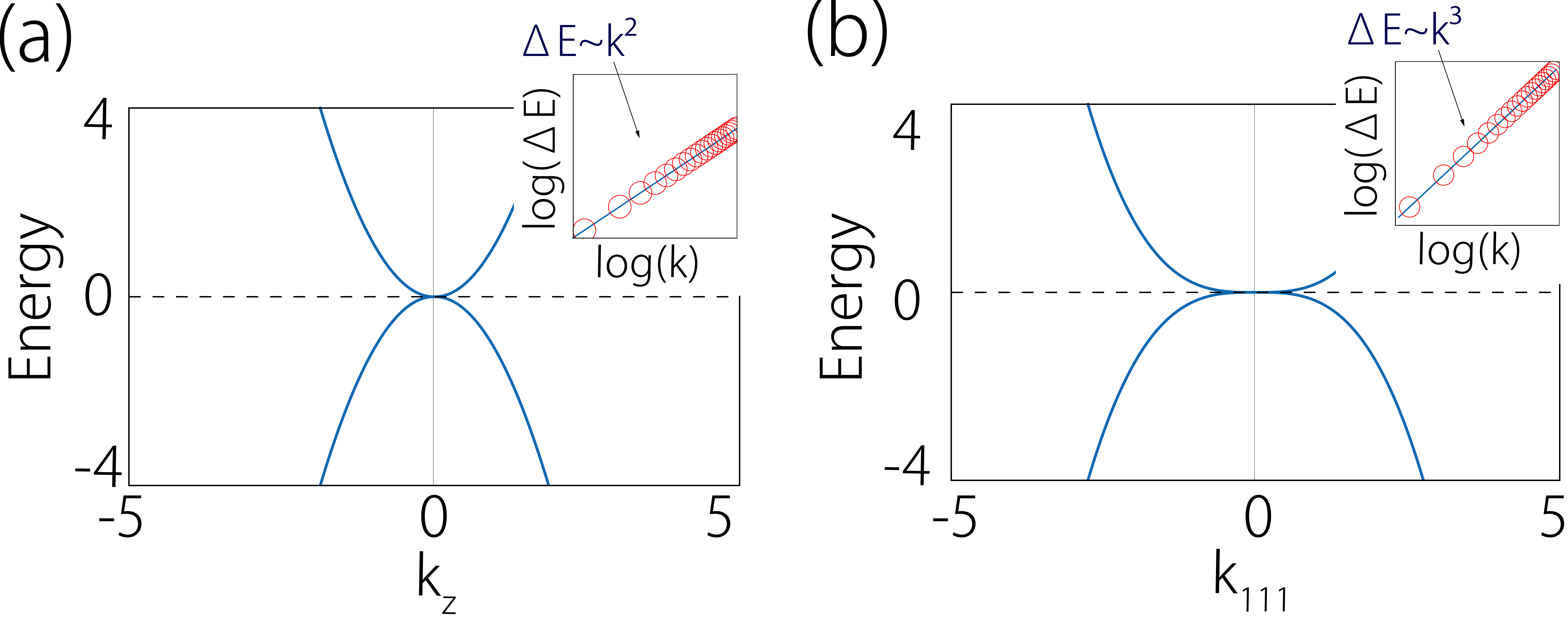}
\par\end{centering}
\caption{The band structure of C-4 WP obtained from the Hamiltonian (\ref{eq:ham1})
with $c_{1}=0$, $c_{2}=1$, $c_3=\frac{1}{3}$ and $c_{4}=\frac{1}{3}$. (a) and (b) show the energy dispersion along $k_{z}$ and $k_{111}$ direction. The insets
are the log-log plot for the band splitting regarding to the momentum, showing the quadratic (a) and cubic (b) energy dispersion.
\label{fig:kpband}}
\end{figure}

The generating elements of PG $T$ can be chosen as a three-fold rotation axis along
$(111)$ direction ($C_{3,111}^{+}$), and two orthogonal two-fold
rotation axis, such as $C_{2x}$ and $C_{2y}$. However, the PG $T$
does not have a 2D single-valued irreducible representation (IR) and
is not capable to protect a C-4 WP. In the addition of ${\cal T}$
symmetry, a pair of conjugated 1D IRs of $C_{3,111}^{+}$ would be
bound together, leading to a 2D corepresentation. The basis state
of this 2D corepresentation then can be chose as $\Psi=(|c_{3}=e^{i2\pi/3}\rangle,\ |c_{3}=e^{-i2\pi/3}\rangle)$
with $c_{3}$ the eigenvalue of $C_{3,111}^{+}$. Under this basis,
the matrix representation of $C_{3,111}^{+}$ and ${\cal T}$ are
\begin{equation}
D(C_{3,111}^{+})=\cos\frac{2\pi}{3}\sigma_{0}+i\sin\frac{2\pi}{3}\sigma_{z},\ \ D({\cal T})=\sigma_{x}{\cal K},
\end{equation}
with $\sigma_{0}$ the $2\times2$ identity matrix, $\sigma_{i}$
($i=x,y,z$) the Pauli matrix and ${\cal K}$ the complex conjugate
operator. $C_{2x(2y)}$ and $C_{3,111}^{+}$satisfy the following
algebra
\begin{eqnarray}
C_{2x(2y)}^{2}=1,\  & C_{2x}C_{2y}=C_{2y}C_{2x}, & C_{3,111}^{+}C_{2y}=C_{2x}C_{3,111}^{+}.\nonumber \\
\end{eqnarray}
Moreover, as $C_{2x(2y)}$ commutes with ${\cal T}$, the matrix representation
of $C_{2x(2y)}$ can be expressed as
\begin{eqnarray}
D(C_{2x}) & = & D(C_{2y})=\sigma_{0}.
\end{eqnarray}
The effective Hamiltonian expanded around the band crossing with this
2D corepresentation is required to be invariant under the symmetry
constraints, namely,
\begin{eqnarray}
D({\cal{O}})H_{k\cdot p}({\cal{O}}^{-1}\boldsymbol{k})D^{-1}({\cal{O}}) & = & H_{k\cdot p}(\boldsymbol{k}),
\end{eqnarray}
where ${\cal{O}}$ runs over the generating elements of PG $T$ and ${\cal T}$.
According to the constraint of $C_{2x}$
\begin{eqnarray}
\sigma_{0}H_{k\cdot p}(k_{x},-k_{y},-k_{z})\sigma_{0} & = & H_{k\cdot p}(\boldsymbol{k}),
\end{eqnarray}
all the first-order terms in $k_{y(z)}$ must be excluded. Similar
analysis applies for $C_{2y}$, which eliminates the first-order terms
in $k_{x(z)}$. Hence, the leading order in $H_{k\cdot p}$ becomes
$k$ quadratic. To preserve $C_{3,111}^{+}$ symmetry, the Hamiltonian
along $(111)$ direction must take the form of $H_{k\cdot p}(k_{111})=f_{1}(k_{111})\sigma_{0}+f_{2}(k_{111})\sigma_{z}$
with $f_{1,2}$ an arbitrary function. However, the linear term in
$f_{1,2}(k_{111})$ is eliminated by $C_{2x(2y)}$, and the quadratic
term in $f_{2}(k_{111})$ is eliminated by ${\cal T}$, as $\{D({\cal T}),k_{111}^{2}\sigma_{z}\}=0$.
Hence, the effective Hamiltonian, to the leading order, reads
\begin{eqnarray}
H_{T} & = & c_{1}k^{2}\sigma_{0}+c_{2}k_{x}k_{y}k_{z}\sigma_{z}\nonumber \\
 &  & +\sqrt{3}\left[c_{3}\left(k_{x}^{2}-k_{z}^{2}\right)+c_{4}\left(k_{y}^{2}-k_{z}^{2}\right)\right]\sigma_{x}\nonumber \\
 &  & -\left[c_{3}\left(k^{2}-3k_{y}^{2}\right)-c_{4}\left(k^{2}-3k_{x}^{2}\right)\right]\sigma_{y},\label{eq:ham1}
\end{eqnarray}
where $k=\sqrt{k_{x}^{2}+k_{y}^{2}+k_{z}^{2}}$ and $c$'s are real coefficients related to specific materials. This Hamiltonian
(\ref{eq:ham1}) exhibits cubic band splitting along $(111)$ direction
and quadratic splitting for any other directions as shown in Fig.
\ref{fig:kpband}, consist with the above analysis. A direct calculation gives that the Chern number for the point
is ${\cal C}=4\times\text{Sign}(c_{2})$ {[}see Fig. \ref{fig:lattice}(d){]},
indicating the existence of the C-4 WP.

With the addiction of $C_{2,110}$ symmetry, the PG $T$ is transformed
in $O$. In contrast to PG $T$, the PG $O$ alone is enough to produce
a C-4 WP. The PG $O$ has only one 2D IR, for which the basis state
also can be as $\Psi=(|c_{3}=e^{i2\pi/3}\rangle,\ |c_{3}=e^{-i2\pi/3}\rangle)$
with $c_{3}$ the eigenvalue of $C_{3,111}^{+}$. The matrix representation
for $C_{2,110}$ under the basis is
\begin{equation}
D(C_{2,110})=\sigma_{x},
\end{equation}
as $C_{2,110}C_{3,111}^{+}=C_{3,111}^{-}C_{2x}C_{2,110}$. The effective
Hamiltonian should be invariant under PG $O$ symmetries, and is obtained
as,
\begin{eqnarray}
H_{O}& = & c_{1}k^{2}\sigma_{0}+c_{2}k_{x}k_{y}k_{z}\sigma_{z}\nonumber \\
 &  & +c_{3}\left[\left(k^{2}-3k_{z}^{2}\right)\sigma_{x}+\sqrt{3}\left(k_{y}^{2}-k_{x}^{2}\right)\sigma_{y}\right],\label{eq:hamkp}
\end{eqnarray}
which shares same band splitting and topological charge with that of Hamiltonian $H_{T}$ (\ref{eq:ham1}). If we further add ${\cal T}$
symmetry to PG $O$, the effective Hamiltonian for the point would
not be changed, as ${\cal T}$ commutes with $H_{O}$ (\ref{eq:hamkp}) and can not further simplifies Eq. (\ref{eq:hamkp}).

\subsection{Minimum lattice model}

In Ref. \cite{yu2021encyclopedia}, a six-band lattice model for the C-4 WP is constructed for the purpose of illustrating the usage of the encyclopedia established
there. However, for the study of C-4 WP, a simple lattice model is favored. According to the relationship between site-symmetry of
Wyckoff position and the elementary band representation \cite{bradlyn2017topological,JenniferPRB2018},
%cite relevant works by Bernevig Nature
we find the minimum lattice model for C-4 WP is a two-band one.

\begin{figure}
\begin{centering}
\includegraphics[width=8cm,height=8cm,keepaspectratio]{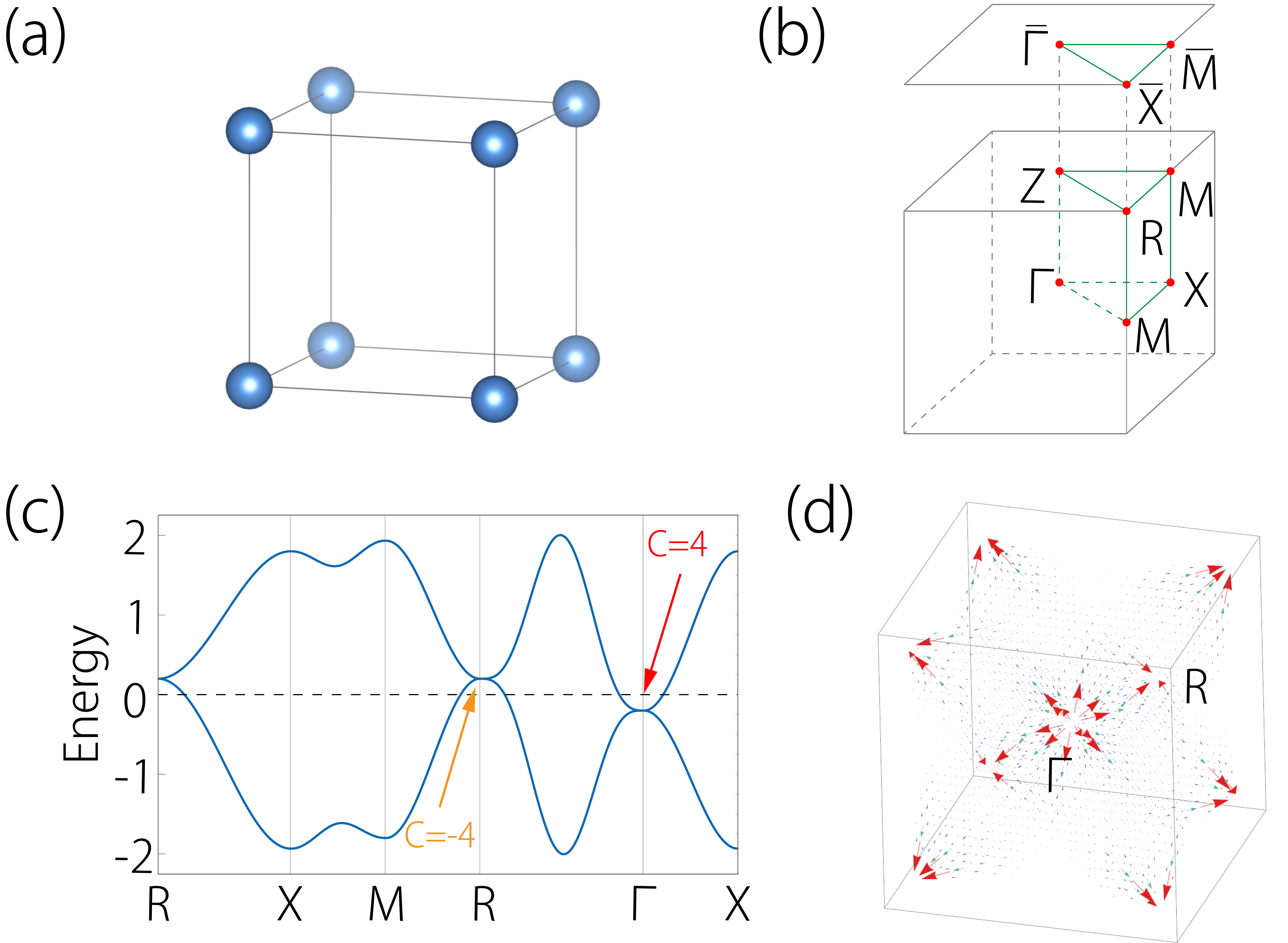}
\par\end{centering}
\caption{(a) Crystal structure for the lattice model of C-4 WP. (b) Bullk and
surface BZs of the lattice. (c) Band structure of the  lattice model (\ref{eq:TB}).
(d) Distribution of Berry curvature in the BZ. In the calculation,
we set $t_{1}=-0.133$, $t_{2}=4$ and $t_{3}=0.539$.\label{fig:lattice}}

\end{figure}

Consider a cubic lattice belonging to SG 207, which is symmorphic SG with PG $O$.
As illustrated in Fig. \ref{fig:lattice}(a), each unit cell contains
an active site locating at $1a$ Wyckoff position. At each site, we
put two basis orbitals $(|d_{x^{2}-y^{2}}\rangle,|d_{z^{2}}\rangle)$
without spin degree of freedom. Under these basis orbitals, the matrix
representation of the generating operators of SG 207 read
\begin{eqnarray}
D(\mathcal{C}_{3,111})=\frac{-\sigma_{0}+i\sqrt{3}\sigma_{y}}{2}, & \  & D(\mathcal{C}_{2x})=D(\mathcal{C}_{2z})=\sigma_{0},\\
D(\mathcal{C}_{2,110})=-\sigma_{z}, & \  & D(\mathcal{T})=\sigma_{0}{\cal K}.
\end{eqnarray}
Following the standard approach \cite{wieder2016spin,yu2019circumventing},
we construct the required lattice model, which in momentum space may be written as
\begin{eqnarray}
H_{\text{TB}} & = & t_{1}(\cos k_{x}+\cos k_{y}+\cos k_{z})\nonumber \\
 &  & +t_{2}\sin k_{x}\sin k_{y}\sin k_{z}\sigma_{y}+3t_{3}(\cos k_{x}-\cos k_{y})\sigma_{x}\nonumber \\
 &  & -\sqrt{3}t_{3}\left(\cos k_{x}+\cos k_{y}-2\cos k_{z}\right)\sigma_{z},\label{eq:TB}
\end{eqnarray}
where $t_{i}$ ($i=1,2,3$) is real model parameters. One can check
the lattice model (\ref{eq:TB}) is invariant under the above symmetry
operators. Also it is easy to find that the two bands of the model
would be degenerate at $\Gamma$ point and $R$ point, and the effective
Hamiltonian expanded around these two points recover the $k\cdot p$
Hamiltonian of C-4 WP (\ref{eq:hamkp}) but with opposite parameters. Hence, the lattice
model contains two C-4 WPs, which respectively locate at $\Gamma$
point and $R$ point and have opposite chirality.

The calculated band structure of this minimum lattice model (\ref{eq:TB})
is shown in Fig. \ref{fig:lattice}(c), where two C-4 WPs locating
at $\Gamma$ point and $R$ point can be observed. Remarkably, these
two WPs are the only two band degeneracies in the model. For comparison,
in the lattice model of Weyl semimetal in spinful systems with ${\cal T}$
symmetry, the number of the WPs must be at least eight, due to the
existence of eight ${\cal T}$-invariant high-symmetry points. The
Chern number of the points in $\Gamma$ point and $R$ point are calculated as $+4$
and $-4$, respectively. To further show the topological configuration,
we plot the Berry curvature ${\cal B}(\boldsymbol{k})$ for the
valence band in Fig. \ref{fig:lattice}(d). The Berry flux emits from the C-4 WP at $\Gamma$ point and converges onto the C-4 WP at $R$
point, consistent with the Chern number of the two WPs.

\section{Chirality-dependent properties }

\subsection{Chiral Landau levels}

The chiral band degeneracies are expected to exhibit chiral Landau
levels (LLs), which can lead to many intriguing phenomena, such as
chiral anomaly and negative longitudinal magnetoresistance \cite{son2013chiral,xiong2015evidence}.
%cite relevant works
Recently, an excellent work by Zhao and Yang \cite{zhao2021index} proved that the number of the chiral
LLs is exactly equal to the topological charge of the degeneracy,
solving the long-term speculation about the connection between these
two properties. Thus, the C-4 WP would always have four chiral LLs
regardless of the direction of magnetic field. In Ref. \cite{yu2021encyclopedia}, Yu et. al. numerically calculated the LL spectrum of C-4 WP when the $B$-field
is along $z$-direction, and indeed observed four chiral LLs crossing the zero energy.

However, for the systems with C-4 WP, the principal axis is $(111)$-axis and it is instructive to study the LL spectrum
with $B$-field along $(111)$-direction. We first rotate the coordinate axis of system and rewrite the effective Hamiltonian of the C-4 WP (\ref{eq:hamkp}) as
\begin{eqnarray}
{\cal H}(\boldsymbol{q}) & = & wq^{2}\sigma_{0}-\alpha_{1}q_{z}(2q_{\parallel}^{2}+q^{2}-3q_{z}^{2})\sigma_{3}\nonumber \\
 &  & +\alpha_{2}\left[(q_{-}^{2}+2\sqrt{2}q_{z}q_{+})\sigma_{+}+h.c.\right],\label{eq:kpqham}
\end{eqnarray}
with $w=c_{1}$, $\alpha_{1}=\frac{c_{2}}{6\sqrt{3}}$, $\alpha_{2}=-c_{3}$, $q_{x}=\frac{2k_{z}-k_{x}-k_{y}}{\sqrt{6}}$, $q_{y}=\frac{k_{x}-k_{y}}{\sqrt{2}}$, $q_{z}=\frac{k_{x}+k_{y}+k_{z}}{\sqrt{3}}$, $q=\sqrt{q_x^2+q_y^2+q_z^2}$, $q_{\parallel}=\sqrt{q_x^2+q_y^2}$, $q_{\pm}=q_{x}\pm iq_{y}$ and $\sigma_{\pm}=\dfrac{1}{2}(\sigma_{x}\pm i\sigma_{y})$.
Here, $q_{z}$ is along the original $(111)$ direction. The Chern number of this model (\ref{eq:kpqham}) is solely determined by the sign of $\alpha_{1}$, namely ${\cal C}=4\times\text{Sign}(\alpha_{1})$.

Interestingly, one finds that for $q_{z}=0$ plane, the Hamiltonian (\ref{eq:kpqham}) becomes
\begin{eqnarray}
{\cal H}(q_{z}=0) & = & wq_{\parallel}^{2}\sigma_{0}+\alpha_{2}(q_{-}^{2}\sigma_{+}+h.c.),
\end{eqnarray}
which exactly is the quadratic Dirac model in bilayer graphene without trigonal warpping effect \cite{nilsson2008electronic}.
And for the $q_{z}\neq0$ plane, the Hamiltonian (\ref{eq:kpqham}) becomes a gapped bilayer graphene model with trigonal warpping effect, and the strength of the warpping effect is proposition to $q_{z}$ \cite{nilsson2008electronic}.

\begin{figure}
\begin{centering}
\includegraphics[width=8cm,height=8cm,keepaspectratio]{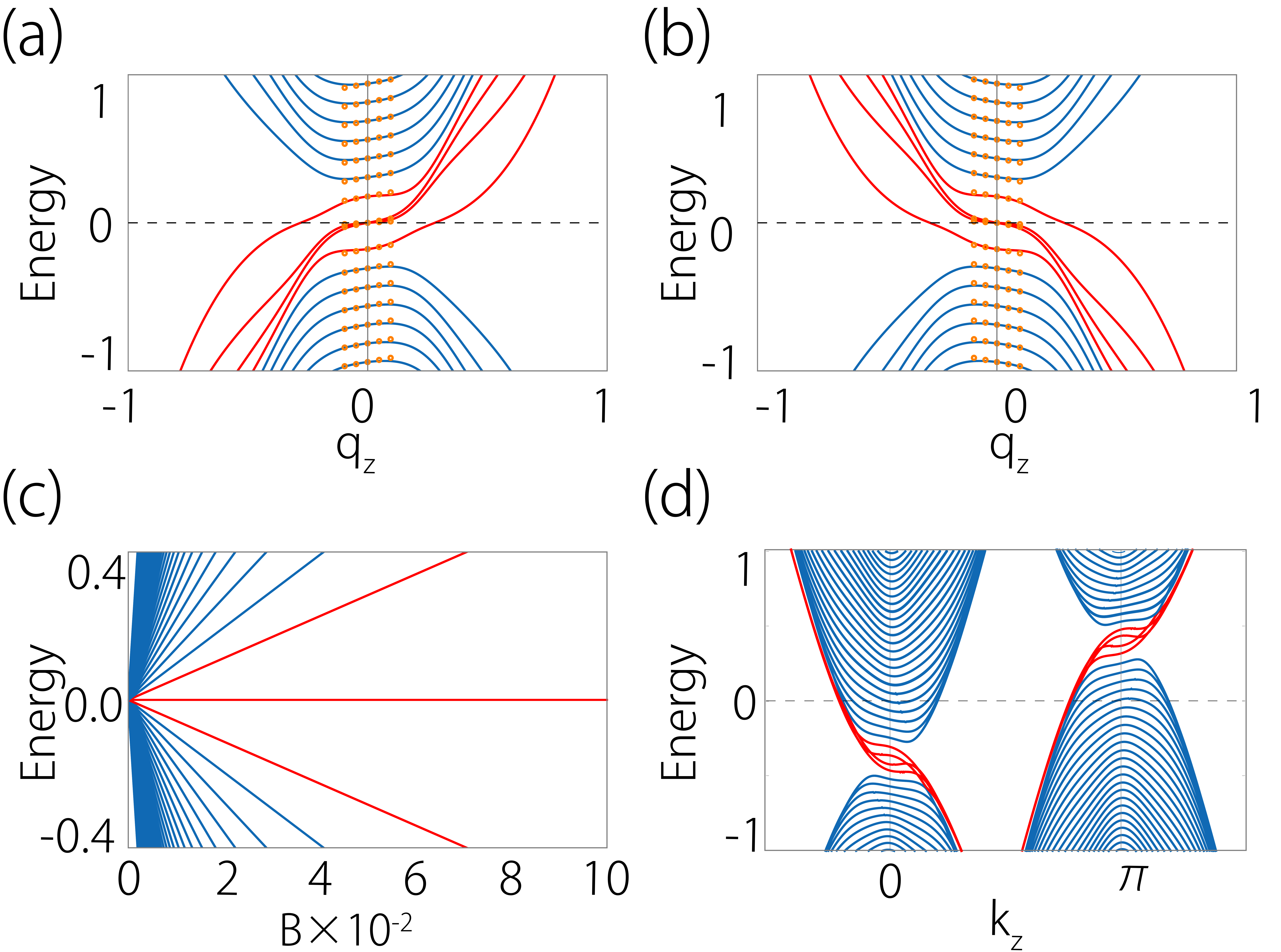}
\par\end{centering}
\caption{(a-c) Landau spectrum calculated from the effective model (\ref{eq:hamkp})
with the $B$-field along $(111)$-direction for (a) $w=0$, $\alpha_{1}=-1$
and $\alpha_{2}=2$, and (b) $w=0$, $\alpha_{1}=1$ and $\alpha_{2}=2$. The orange circles are obtained by first-order perturbation.
(c) shows the Landau spectrum as a function of $B$ for $q_{z}=0$
plane. (d) Landau spectrum calculated from the lattice model (\ref{eq:TB})
with the $B$-field along $z$-direction. \label{fig:LLs}}

\end{figure}

By applying an external magnetic field along $q_{z}$-direction, the
electron's motion in $q_{x}$-$q_{y}$ plane is quantized into LLs.
We apply the usual Peierls substitution $\boldsymbol{q}\rightarrow\boldsymbol{q}+e\boldsymbol{A}$
in Eq. (\ref{eq:kpqham}) with $\boldsymbol{A}$ the vector potential.
The LL spectrum of C-4 WP at $q_{z}=0$ plane can be easily established
as
\begin{eqnarray}
\varepsilon_{n=0}(q_{z}=0)=w B, & \ \  & \varepsilon_{n=1}(q_{z}=0)=3w B,
\end{eqnarray}
for $n=0,1$ and
\begin{eqnarray}
\varepsilon_{n}(q_{z}=0) & = & \left[(2n-1)w\pm2\sqrt{w^{2}+\alpha_{2}^{2}\left(n-1\right)n}\right]B, \nonumber\\
\end{eqnarray}
for $n>1$.
%, with $\ell_{B}=\sqrt{1/(eB)}$.
The LL spectrum has a linear dependence on the magnetic field as shown in Fig. \ref{fig:LLs}(c),
similar to the conventional electron gas model. Moreover, when $w=0$,
the first two LLs ($n=0,1$) would be degenerate at zero energy, and are not
sensitive to the field strength. Due to the presence of trigonal warpping
effect, the analytical expressions of the LL spectrum for $q_{z}\neq0$
plane generally can not be obtained. Since the low-energy physics
of the LLs, which is relevant for experimental observation, is dominated
by the electrons residing at $q_{x}$-$q_{y}$ plane with small $q_{z}$,
one generally can use the perturbation theory to obtain the low-energy LL spectrum
of C-4 WP.
We find that the analytic solutions obtained from the first-order perturbation are in good agreement with the numerical results in the low-energy region, as shown in Fig. \ref{fig:LLs}(a-b).

Particularly, four chiral LLs can be clearly observed from the numerically results of the LLs, and one finds that the slope of the four chiral LLs is determined by the sign of $\alpha_{1}$ [see Fig. \ref{fig:LLs}(a-b)]. To further demonstrate the chirality-dependent properties of the chiral LLs, we show the numerical result of the Landau spectrum based on the lattice model (\ref{eq:TB}) with the magnetic field along $z$-direction. Four chiral LLs appear around
$k_{z}=0$ with negative slope, corresponding to the C-4 WP with ${\cal C}=4$ at $\Gamma$ point {[}see Fig. \ref{fig:lattice}(b){]}. Meanwhile, another four chiral  LLs appear around $k_{z}=\pi$ with opposite slope, induced by the C-4 WP with ${\cal C}=-4$ at $R$ point.

\subsection{Quantized circular photogalvanic effect }

\begin{figure}
\begin{centering}
\includegraphics[width=8cm,height=8cm,keepaspectratio]{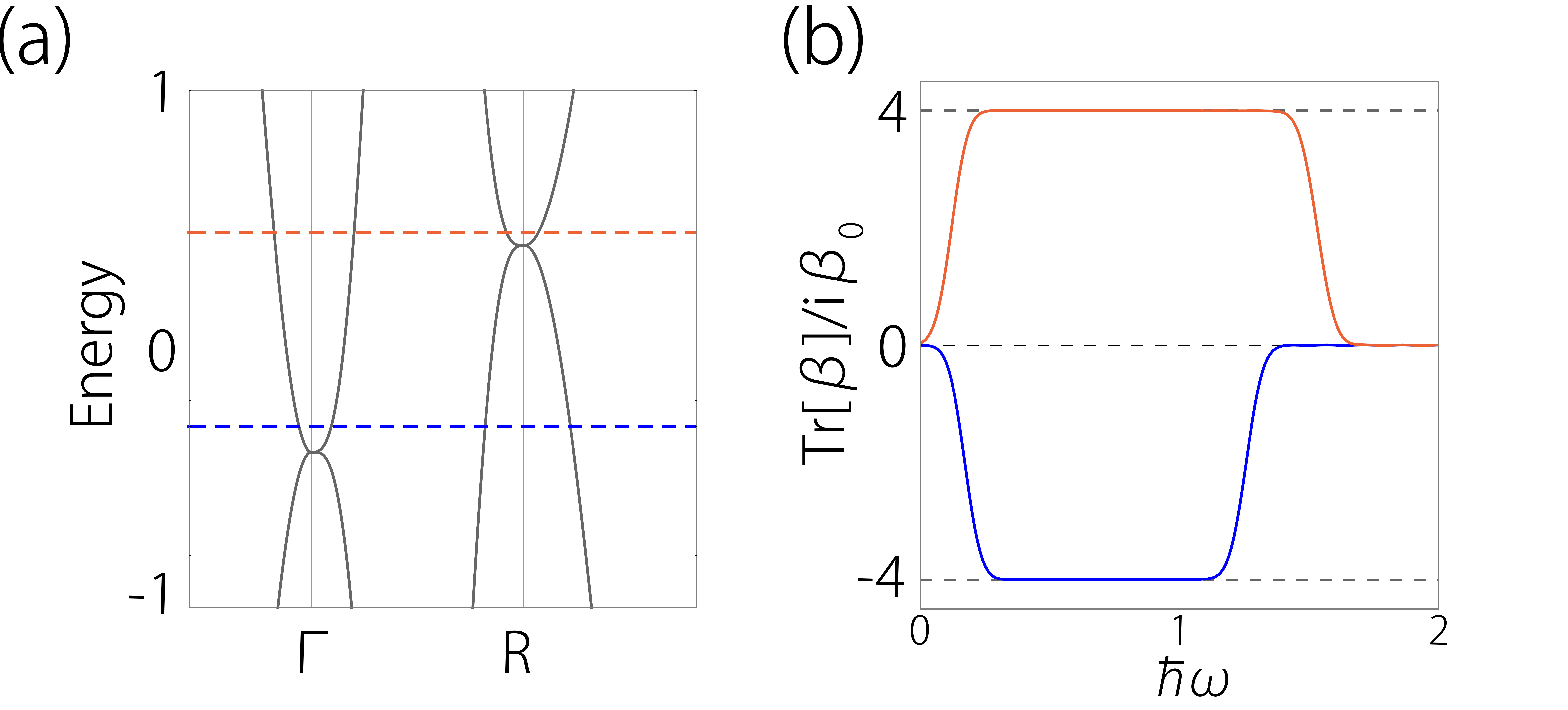}
\par\end{centering}
\caption{(a) The band structure of C-4 WP semimetal calculated from the lattice
model (\ref{eq:TB}). (b) CPGE traces at different chemical potential
($\mu=-0.3$ and $0.45$), denoted as dashed lines in (a).
\label{fig:CPGE}}

\end{figure}

Another interesting chirality-dependent phenomena for chiral particle may be the quantized CPGE \cite{de2017quantized}.
The CPGE injection current is defined as
\begin{equation}
\frac{dj_{i}}{dt}=\beta_{ij}(\omega)\left[\mathbf{E}(\omega)\times\mathbf{E}^{*}(\omega)\right]_{j}
\end{equation}
where $\mathbf{E}(\omega)=\mathbf{E}^{*}(-\omega)$ is the electric
field. The CPGE tensor $\beta_{ij}$ is purely imaginary and can be
calculated by

\begin{equation}
\beta_{ij}(\omega)=\frac{\pi e^{3}}{\hbar V}\epsilon_{jkl}\sum_{\boldsymbol{k},n,m}f_{nm}^{\boldsymbol{k}}\Delta_{\boldsymbol{k},nm}^{i}r_{\boldsymbol{k},nm}^{k}r_{\boldsymbol{k},mn}^{l}\delta(\hbar\omega-\varepsilon_{\boldsymbol{k},mn}),
\end{equation}
with $V$ the sample volume, $\varepsilon_{\boldsymbol{k},mn}=\varepsilon_{n,\boldsymbol{k}}-\varepsilon_{m,\boldsymbol{k}}$
($f_{nm}^{\boldsymbol{k}}=f_{n}^{\boldsymbol{k}}-f_{m}^{\boldsymbol{k}}$)
the energy (Fermi-Dirac distributions) difference between $n$- and
$m$-the bands, $r_{\boldsymbol{k},nm}=i\langle u_{n}|\partial_{\boldsymbol{k}}|u_{m}\rangle$
and $\Delta_{\boldsymbol{k},nm}^{i}=\partial_{k_{i}}\varepsilon_{\boldsymbol{k},nm}$.
Juan et al. \cite{de2017quantized} demonstrated that for WPs, the trace of the CPGE tensor can be quantized    and its values is proportional to the topological
charge ${\cal C}$ of WPs,
\begin{equation}
\text{Tr}(\beta)=\sum_{i=x,y,z}\beta_{ii}=-i\beta_0{\cal C},
\end{equation}
with $\beta_0=\frac{e^{3}}{h^{2}}\pi$.
A direct calculation shows the quantized CPGE also applies for the C-4 WP and the quantized value is $\pm4$ depending on the topological
charge of the C-4 WP. Based on the lattice model (\ref{eq:TB}), we numerically calculate the injection current with different Fermi energy,
as shown in Fig. \ref{fig:CPGE}. Our results show the quantized CPGE indeed can appear in C-4 WP semimetal with suitable Fermi energy and frequency $\omega$. When the CPGE is quantized, the sign of this quantized quantity is determined by the chirality of the C-4 WP, consistent with the theoretical analysis.
For example, when the Fermi energy is $0.45$, then only the C-4 WP at $R$ point has contribution to the CPGE effect for $0.1<\omega<1.7$. In contrast, when the Fermi energy is set as $-0.3$, then only the C-4 WP at $\Gamma$ point is activated for $0.2<\omega<1.4$.

\subsection{Quadruple-helicoid surface states}

Topological Weyl semimetals exhibit novel Fermi arc state on the boundary
of systems. Remarkably, the Fermi arc state around the projected point
of the WP also shows chirality-dependent properties. Since the C-4
WP has a topological charge of $|{\cal C}|=4$, four Fermi arcs would
be observed in C-4 WP semimetal. One striking feature of the Fermi arc surface state is that its isoenergy contour is not closed but consist of one or multiple open curves. Fang et al. \cite{fang2016topological} for the first time pointed out that the surface state of a conventional
C-1 WP around the projected point on boundary is equivalent to a helicoid.
Later, Zhang et al. \cite{zhang2018double} showed the surface state of a Dirac point with
topological charge of $|{\cal C}|=2$ is a double helicoid. Thus,
one can expect the surface state around the projection of a C-4 WP would be a quadruple helicoid, namely, four screw surfaces wind around the C-4 WP.

\begin{figure}
\begin{centering}
\includegraphics[width=8cm,height=8cm,keepaspectratio]{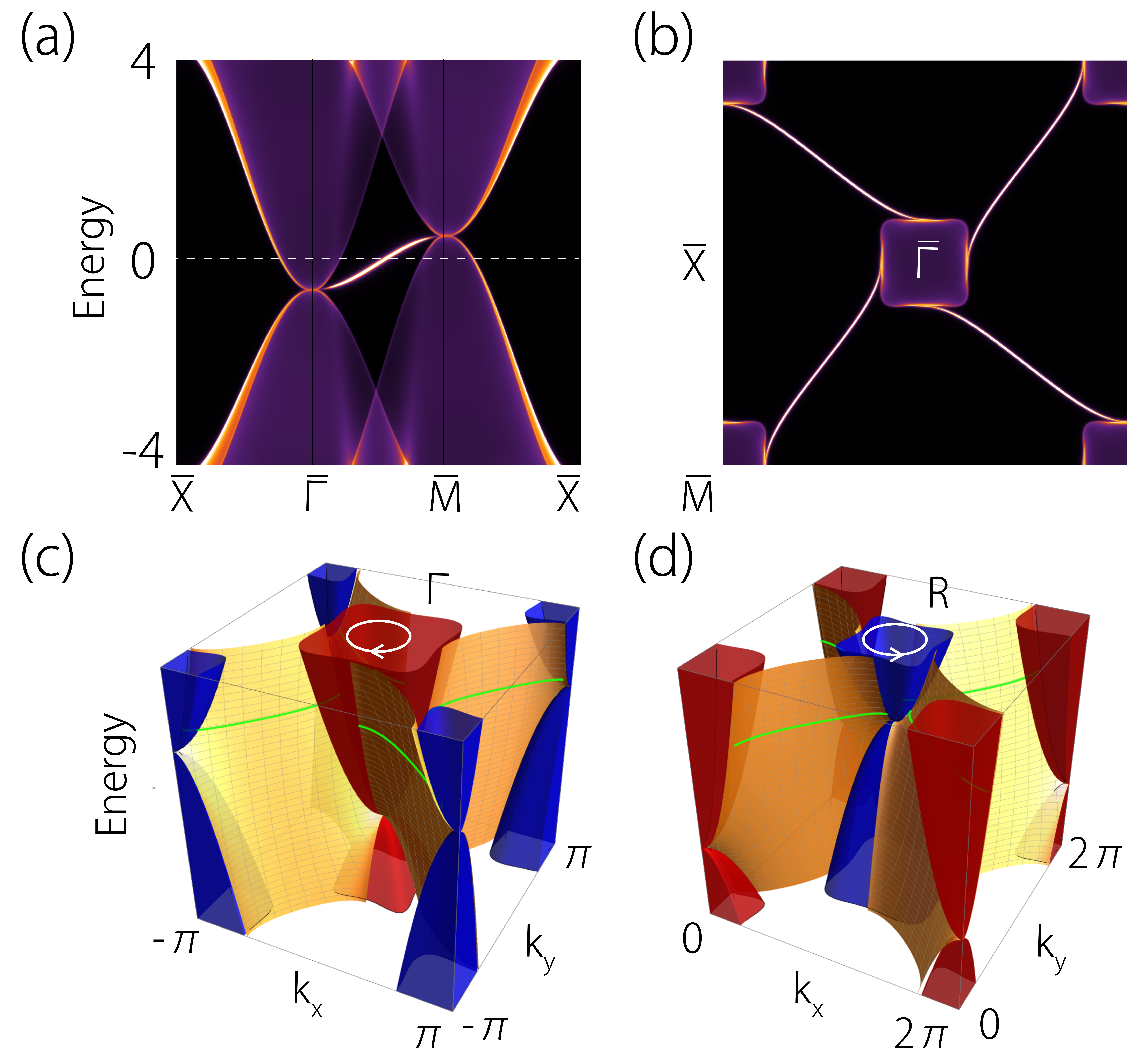}
\par\end{centering}
\caption{Surface spectra of the lattice model in Fig. \ref{fig:lattice}(c).
(a) Projected spectrum on (001) surface. (b) The slice at constant
energy $E_{F}=0$ on (001) surface. (c-d) The quadruple helicoid surface
state calculated from the lattice model (\ref{eq:TB}) confined in
$z$-direction. (c) and (d) show $(001)$ surface bands centered at
$\bar{\Gamma}$ and $\bar{M}$, respectively. Red and blue
surfaces denote bulk band spectra, and the orange surfaces denote
the four arc surface states. Green line show four Fermi arcs at a
given energy. \label{fig:arc}}

\end{figure}

In Fig. \ref{fig:arc}, we calculated the $(001)$ surface state of the lattice model (\ref{eq:TB}). The two C-4 WPs are projected to $\bar{\Gamma}$
and $\bar{M}$ points of the surface BZ. Four extensive Fermi arcs
connecting the $\bar{\Gamma}$ and $\bar{M}$ points can be observed,
as shown  in Fig. \ref{fig:arc}(b). To further show the geometry
structure of the surface state, we calculate the band structure of
a slab (confined in $z$-direction) based on the model (\ref{eq:TB}). The obtained results  are plotted in Fig. \ref{fig:arc}(c) and \ref{fig:arc}(d). It indeed shows four screw surfaces, e.g. the quadruple-helicoid surface state around $\bar{\Gamma}$ and $\bar{M}$ points. Particularly, the chirality of the
helicoid reflects the chirality of the C-4 WPs. For example, by increasing energy, the four Fermi arcs clockwise wind around $\bar{\Gamma}$
point, while they anti-clockwise wind around $\bar{M}$ point. Hence, the helicoids around $\bar{\Gamma}$ point and  $\bar{M}$ point have opposite chirality, consistent with chirality of the two C-4 WPs at $\Gamma$ and $R$ points.

\section{Discussion and Conclusion }

\begin{figure}
\begin{centering}
\includegraphics[width=8cm,height=8cm,keepaspectratio]{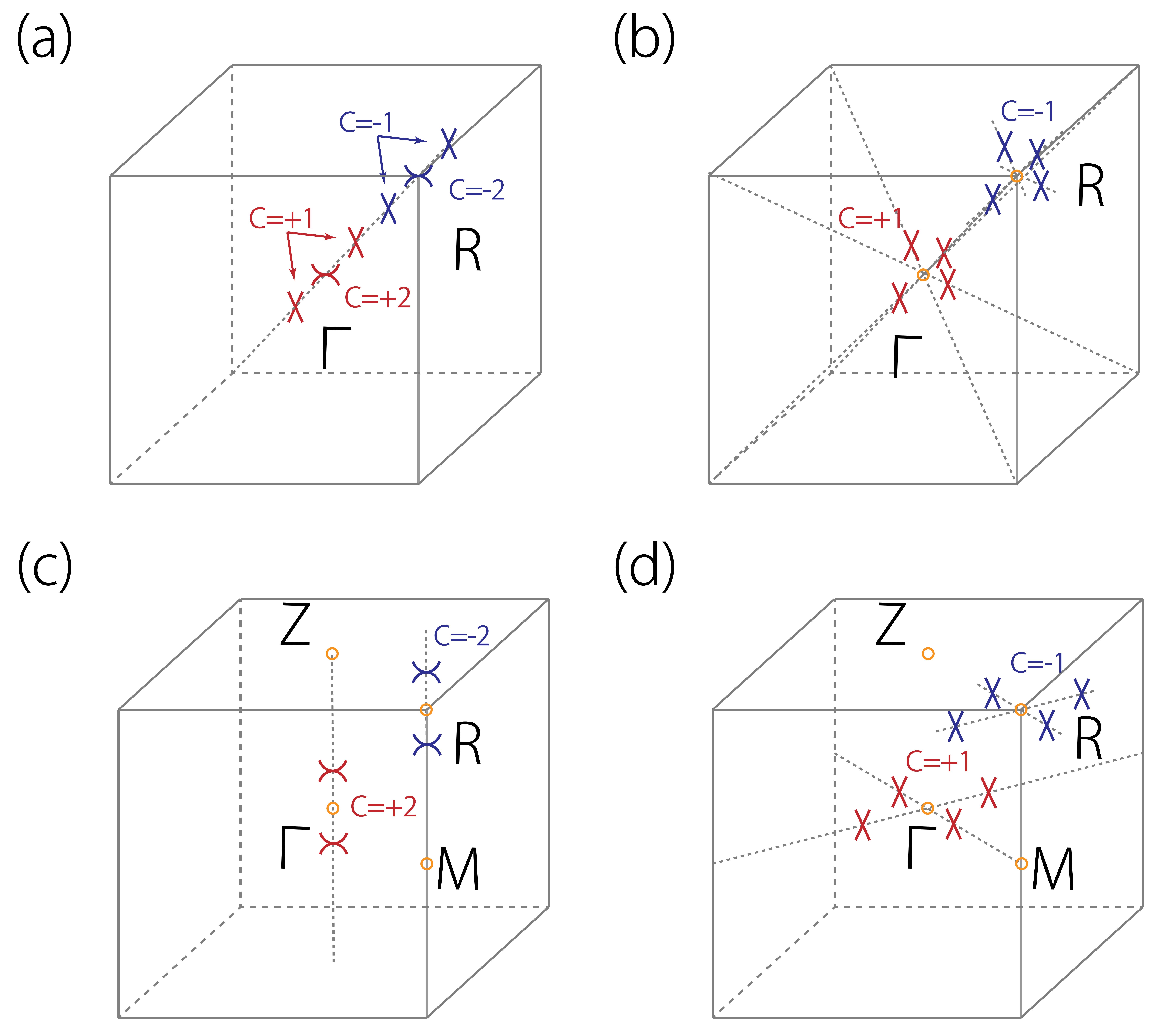}
\par\end{centering}
\caption{C-4 WP under symmetry breaking. Start from the lattice model in Fig.
\ref{fig:lattice}(c). (a) breaks $C_{2,110}$ and $C_{2x}$ symmetry,
(b) breaks $C_{2,110}$ and ${\cal T}$ symmetry and (c-d) only breaks
$C_{3,111}^{+}$ symmetry. \label{fig:sysbreaking }}
\end{figure}

Since the C-4 WP is protected by symmetry, it is interesting  to study its evolution under symmetry breaking. Several interesting cases are
shown in Fig. \ref{fig:sysbreaking }. Start from the lattice model
(\ref{eq:TB}), which has PG $O$ and ${\cal T}$ symmetry. Breaking
$C_{2,110}$, the PG $O$ descends to PG $T$, the two C-4 WPs at
$\Gamma$ and $R$ points will persist. It again demonstrates that
the PG $T$ together with ${\cal T}$ symmetry can stabilize the C-4
WP. By further breaking $C_{2x}$, the C-4 WP at $\Gamma$ ($R$)
point would splitting into a C-2 WP locating at the original position
and two C-1 WP residing on $(111)$ axis, as shown in Fig. \ref{fig:sysbreaking }(a).
If we break ${\cal T}$ instead of $C_{2x}$, then the C-4
WP would be transformed into four C-1 WPs, which occur on the body
diagonals of BZ, as shown in Fig. \ref{fig:sysbreaking }(b). Moreover,
when only $C_{3,111}^{+}$ is broken, the C-4 WPs may split into two
C-2 WPs appearing at $\Gamma$-$Z$ ($R$-$M$) path {[}Fig. \ref{fig:sysbreaking }(c){]}
or four C-1 WPs residing at $\Gamma$-$M$ ($Z$-$R$) path {[}Fig.
\ref{fig:sysbreaking }(d){]}, depending on model parameters.

Since the C-4 WP only appears in spinless systems, it cannot be realized in real materials with sizable SOC effect. Nevertheless, the C-4 WP may occur at phonon spectrum, as shown in Ref. \cite{zhang2020twofold}, and the materials with negligible SOC effect. More intriguing possibility may be the artificial crystals, such as the photonic and acoustic crystals, which are spinless systems, and can be well controlled by current technology \cite{lu2013weyl,chen2016photonic,li2018weyl,he2018topological}.
%cite 3-4 recent works published on PRL, Nature, Science
Hence, one can expect that the C-4 WP can be realized in these artificial systems, and the chirality-dependent properties of C-4 WP predicted here then would be  readily for detection.

In conclusion, we systematically study the symmetry requirement, and the electronic, optical and magnetic properties of a newly found emergent particle, e.g. the C-4 WP. We show the minimum symmetry requirement for stablizing a C-4 WP is PG $O$ or PG $T$ together with ${\cal T}$ symmetry. We also construct a minimum lattice model for the C-4 WP. Based on both the low-energy effective model and minimum lattice model, we investigate the Landau spectrum, quantized CPGE, and quadruple-helicoid surface states of the C-4 WP, and show they all have a strong dependence on the chirality of the C-4 WP. At last, we show under symmetry breaking, the C-4 WP would be transformed into many different topological phases.

\begin{acknowledgments}
  The authors thank J. Xun for helpful discussions. This work is supported by the National Key R\&D Program of China (Grants No. 2020YFA0308800), the NSF of China (Grants No. 12061131002, No. 12004035, and No. 11734003), the Strategic Priority Research Program of the Chinese Academy of Sciences (Grant No. XDB30000000), and the Beijing Institute of Technology Research Fund Program for Young Scholars.
\end{acknowledgments}

\bibliography{reference}

\end{document}